\documentclass[namedreferences]{SolarPhysics}
\usepackage[optionalrh]{spr-sola-addons} % For Solar Physics 
\usepackage{graphicx}        % For eps figures, newer & more powerfull
\usepackage{color}           % For color text: \color command
\usepackage{url}             % For breaking URLs easily trough lines
\usepackage{soul}
\begin{document}

\begin{article}

\begin{opening}

\title{Comparisons of CME morphological characteristics derived from five 3D reconstruction methods  }

\author{L.~\surname{Feng}$^{1,2}$\sep
        B.~\surname{Inhester}$^{2}$\sep
        M.~\surname{Mierla}$^{3,4}$ }

\runningauthor{L. Feng}
\runningtitle{Comparisons of 3D reconstructions of CMEs}

\institute{$^{1}$ Key Laboratory of Dark Matter and Space Astronomy, 
           Purple Mountain Observatory, Chinese Academy of Sciences, 
           210008 Nanjing, China
                   email: \url{lfeng@pmo.ac.cn}\\         
           $^{2}$ Max-Planck-Institut f\"{u}r Sonnensystemforschung, Max-Planck-Str.2,
             37191 Katlenburg-Lindau, Germany
                     email: \url{binhest@mps.mpg.de} \\
           $^{3}$ Institute of Geodynamics of the Romanian Academy
                  Jean-Louis Calderon 19-21, Bucharest-37, Romania, RO-020032
                     email: \url{marilena@geodin.ro}\\
           $^{4}$ Royal Observatory of Belgium, Avenue Circulaire 3, 1180 Brussels, Belgium\\}

\begin{abstract}

We compare different methods to reconstruct the three-dimensional (3D) 
CME morphology. The explored
methods include geometric localisation, mask fitting, forward modeling, 
polarisation ratio and local correlation tracking plus triangulation.
The five methods are applied to the same CME event, which occurred on August 7 2010.
Their corresponding results are presented and compared, especially in their
propagation direction and spatial extent in 3D. 
We find that mask fitting and geometric localisation method produce consistent
results. Reconstructions including three-view observations are more precise
than reconstructions done with only two views. Compared to the forward 
modeling method, in which a-priori shape of the CME geometry is assumed,
mask fitting has more flexibility. Polarisation ratio method makes use of the
Thomson scattering geometry. We find spatially the 3D CME derived from mask fitting 
lies mostly in the overlap region obtained with the polarisation method from COR2 A
and B. In addition, mask fitting can help resolve the front/back ambiguity inherent in the 
polarisation ratio method. However, local correlation tracking plus triangulation
did not show a consistent result with the other four methods. For reconstructions
of a diffuse CME, when the separation angle between STEREO A and B is large,
finding two corresponding points in a STEREO image pair becomes very difficult. Excluding the
local correlation tracking method, the latitude of the CME's centre of gravity 
derived from the other methods deviates within one degree and longitude differs within 
19 degrees.          
\end{abstract}


\end{opening}
%-------------------------------------------------

\section{Introduction}

Coronal mass ejections (CMEs) are violent eruptions from the Sun and 
also known as the main cause of major geomagnetic storms. They 
can now be observed from three viewpoints almost simultaneously 
using the two separated Solar Terrestrial Relations
Observatory (STEREO) spacecraft \cite{Kaiser:etal:2008} and SOlar 
and Heliospheric Observatory (SOHO) \cite{Domingo:etal:1995}. COR 1 \& 2 are white-light 
coronagraphs in the Sun Earth Connection Coronal and Heliospheric 
Investigation (SECCHI; \inlinecite{Howard:etal:2008}) instrument package onboard STEREO. 
Large and Spectroscopic Coronagraph (LASCO; \inlinecite{Brueckner:etal:1995}) 
C2 and C3 are white-light coronagraphs onboard SOHO. They provide time-resolved polarised and total brightness images 
of eruptions in the corona. Such measurements can be used to derive the 
location and 3D structure of CMEs, which is crucial for understanding the origin 
and dynamics of these eruptions and for predicting their
effects on the Earth's magnetosphere. 

A number of methods have been developed to obtain the 3D morphology 
of CMEs from multi-view coronagraph data. Mierla {\it et al.} (2009, 2010)
and \inlinecite{Thernisien:etal:2011}
compared the results of different reconstruction methods.
\inlinecite{deKoning:etal:2009} calculated a stack of quadrilaterals which
contain the 3D CME position and shape. \inlinecite{Byrne:etal:2010} fitted an ecllipse 
to the quadrilateral in an attempt to derive a smooth CME shape. 
A different approach to estimate the 3D CME morphology uses forward modelling
assuming a graduated cylindrical shell flux rope model for the CME
\cite{Thernisien:etal:2009,Thernisien:2011}. This flux rope model
is controlled by a few free parameters determined by fitting the flux rope with 
the observed CME from different viewpoints. 
Another method, the polarisation method, was first
applied by \inlinecite{Moran:Davila:2004}. It makes use of the different
anisotropy of the differential Thomson scatter cross section for
different polarisations of the scattered photon. This effect allows
to estimate a virtual centre of the scatterd signal along each
line-of-sight from the observed polarisation ratio.

\inlinecite{Feng:etal:2012} developed a new mask fitting method to achieve
the CME localisation and morphology. It has been used to determine the CME's propagation 
direction and interpret the in-situ observations from different viewpoints. 
The morphological evolution of the CME was analysed as well. The orientation of the 
neutral line in the source region was compared with the major direction of 
the reconstructed 3D cloud and the flux rope axis orientation from the in-situ data.
To validate the newly developed reconstruction method and to seek a reliable method
for the purpose of space weather prediction, in this paper
the mask fitting method is compared with a few other methods which have been used to reconstruct
the 3D shape. In \S 2 the observations of a specific CME is presented. 
Different methods applied to the same CME event are introduced and their corresponding
results can be found in \S 3.  In \S 4 we compare and combine the mask fitting method with 
the other four methods. The final section presents the discussions and conclusions.

\section{Observations and data reduction}

\begin{table} 
\caption{Longitude and latitude of STEREO A \& B, and SOHO in
Carrington coordinate system.}
\begin{tabular}{cccc}
spacecraft & B & SOHO & A\\
\hline
longitude & 311.49 & 22.86 & 102.21\\
latitude & -1.83 & 6.19 & 4.81\\
separation with SOHO & 71.61 & &79.01\\
separation(A\&B) & & 150.61&\\ 
\hline
\end{tabular}    
\label{tab:spa_pos}
\end{table}

\begin{figure} % fig 1
  \centering
  \includegraphics[bb=45 20 125 120, clip=true, height=0.25\textheight]
                   {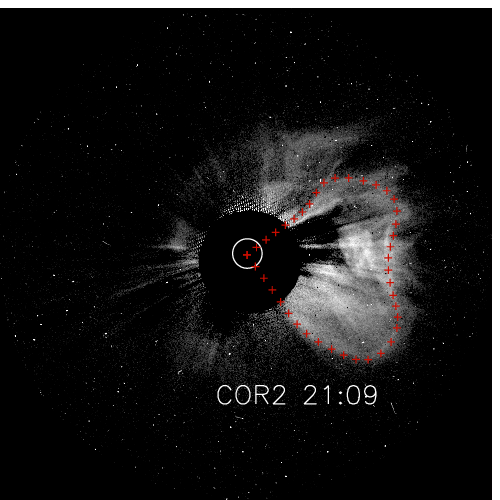}
  \includegraphics[bb=30 40 78 100, clip=true, height=0.25\textheight]
                   {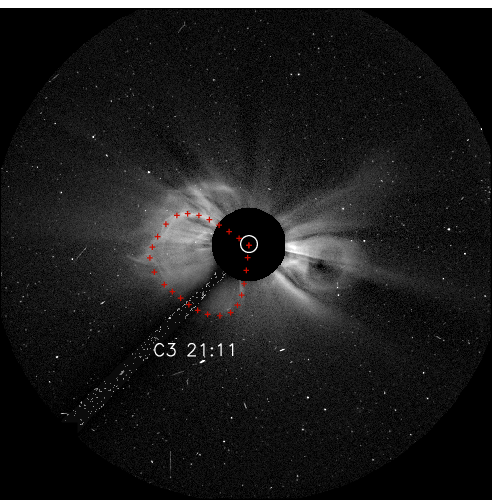}
  \includegraphics[bb=2 20 82 120, clip=true, height=0.25\textheight]
                   {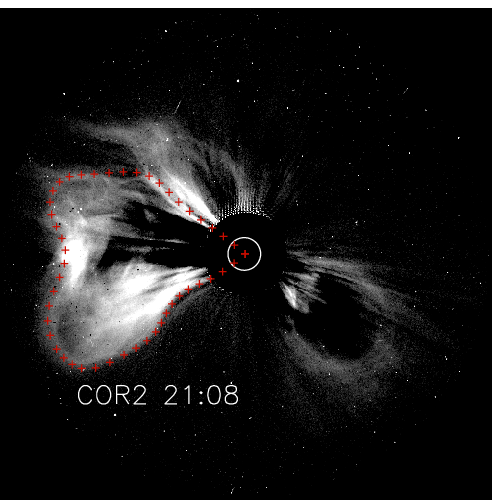}
  \caption{The corograph images recorded by SECCHI/COR2 B at 21:09:05, 
   by LASCO/C3 at 21:11:18 and by SECCHI/COR2 A at 21:08:15. The red 
   curves outline the CME region traced by hand.}
  \label{fig:cor_c3_2108}
\end{figure}

The CME for which the 3D morphology is reconstructed was observed on August 7 2010.
Its propagation in interplanetary space can be traced in the field of view of 
COR and HI in the SECCHI instrument package onboard STEREO.
The CME of interest was observed as well by LASCO C2 and C3 onboard SOHO. In this paper,
the 3D CME cloud is reconstructed only from Sun-centred coronagraph images. 
We are not dealing with the data observed by the Heliospheric Imager at this stage. The 
data from SECCHI/COR2 and 
LASCO/C3 are used. More generally, the methods introduced in the following section 
can be applied to any other coronagraph data if it includes polarisation measurements.

SECCHI/COR2 has a field of view extending from 2.5 to 15 $R_{\odot}$, LASCO/C3 has a larger field 
of view extending from 3.7 to 30 $R_{\odot}$. The spatial resolution of COR2 and C3 are
15 and 56 arcsec/pixel, respectively. COR2 and C3 both have a polariser which 
allows measurements of total, polarised and unpolarised brightness. The total time required
for a polarimetric observation sequence for COR2 is around 20 s, while the corresponding
observation using LASCO is around 300 s. The shorter exposure times eliminate errors
arising from CME motion between exposures.
All the processing of coronagraph images are done by $secchi\_prep$ and 
$lasco\_prep$ in SolarSoft (SSW). For a 3D reconstruction from images observed at two 
or three viewpoints, the positions of each spacecraft are required. In 
Table~\ref{tab:spa_pos} we list the longitude and latitude of STEREO A \& B, and SOHO 
in the Carrington coordinate system. The separation angles between spacecraft are also mentioned.

In Figure~\ref{fig:cor_c3_2108} the background subtracted coronagraph images
are presented. To make the CME signature more prominent, the background for COR2 is a 
pre-CME image recorded about 40 minutes
before the eruption. Its subtraction can remove the stray light and non-CME
coronal emissions. The background for C3 is a computed 12-hour minimal image centred in time at
the CME eruption. Such background subtraction mainly removes the stray light 
and the streamer is still visible.
The CME periphery was traced by hand and fitted by a parametric cubic spline.
The CME-edge tie-points are indicated by red crosses for a time instance at around
21:10 UT. In our calculations we ignored the three-minute difference between the observations of
COR2 A \& B and C3. We have considered the interpolation of the C3 CME from two 
neighbouring frames in time. However, we found that the observational time of one 
of the neighbouring frames was almost half an hour earlier. It is very probable that the 
error induced by the interpolation would exceed the error from the CME propagation
in three minutes. In Figure~\ref{fig:cor_c3_2108} shock signatures were observed 
which are not included in the calculations below. 

The most prominent activity on the solar disk occurred at the time of the CME launch 
in the active region AR~11093 located at N12E31 as viewed from Earth 
on August 7. According to 
GOES light curves, a M-class flare occurred in this active region. 
It started around 17:55 UT and peaked at 18:24 UT. AR~11093 was visible
in the east with respect to the central meridian as seen from the Earth 
observed by the Atmospheric Imaging Assembly (AIA; \inlinecite{Lemen:etal:2011})
onboard the Solar Dynamic Observatory (SDO), whereas in the west in 
Extreme Ultraviolet Imager (EUVI; \inlinecite{Wuelser:etal:2004})
as seen from STEREO B. It was behind the solar limb as
seen from STEREO A. More details of the investigated
CME and its source region can be found in \inlinecite{Feng:etal:2012}.

\section{Descriptions and results of different 3D reconstruction methods}

For direct comparisons of different reconstruction methods, all the calculations
are done for the same CME event and in the same Cartisian Carrington coordinate system.
$X$ and $Y$
axes lie in the solar equatorial plane, where $X$ axis points to zero  
Carrington longitude. $Z$ axis is the solar rotation axis.

\subsection{Geometric localisation (GL)}

\begin{figure} %fig 2
  \centering
  %\setlength{\unitlength}{1cm}
  %\begin{picture}(12,12)
  %\put(0.25,6.5){\includegraphics[width=0.45\textwidth,clip=]{fig2a.eps}}
  %\put(5.8,6.75){\includegraphics[width=0.4\textwidth,clip=]{fig2b.eps}}
  %\put(0.2,0.5){\includegraphics[width=0.425\textwidth,clip=]{fig2c.eps}}
  %\put(6.25,0.0){\includegraphics[width=0.45\textwidth,clip=]{fig2d.eps}}
  %\put(0.,12){\makebox(0,0){a}}
  %\put(5.75,12){\makebox(0,0){b}}
  %\put(0.,5){\makebox(0,0){c}}
  %\put(6.,5){\makebox(0,0){d}}
  %\end{picture}
  \vbox{
  \includegraphics[width=12cm, height=6.cm]{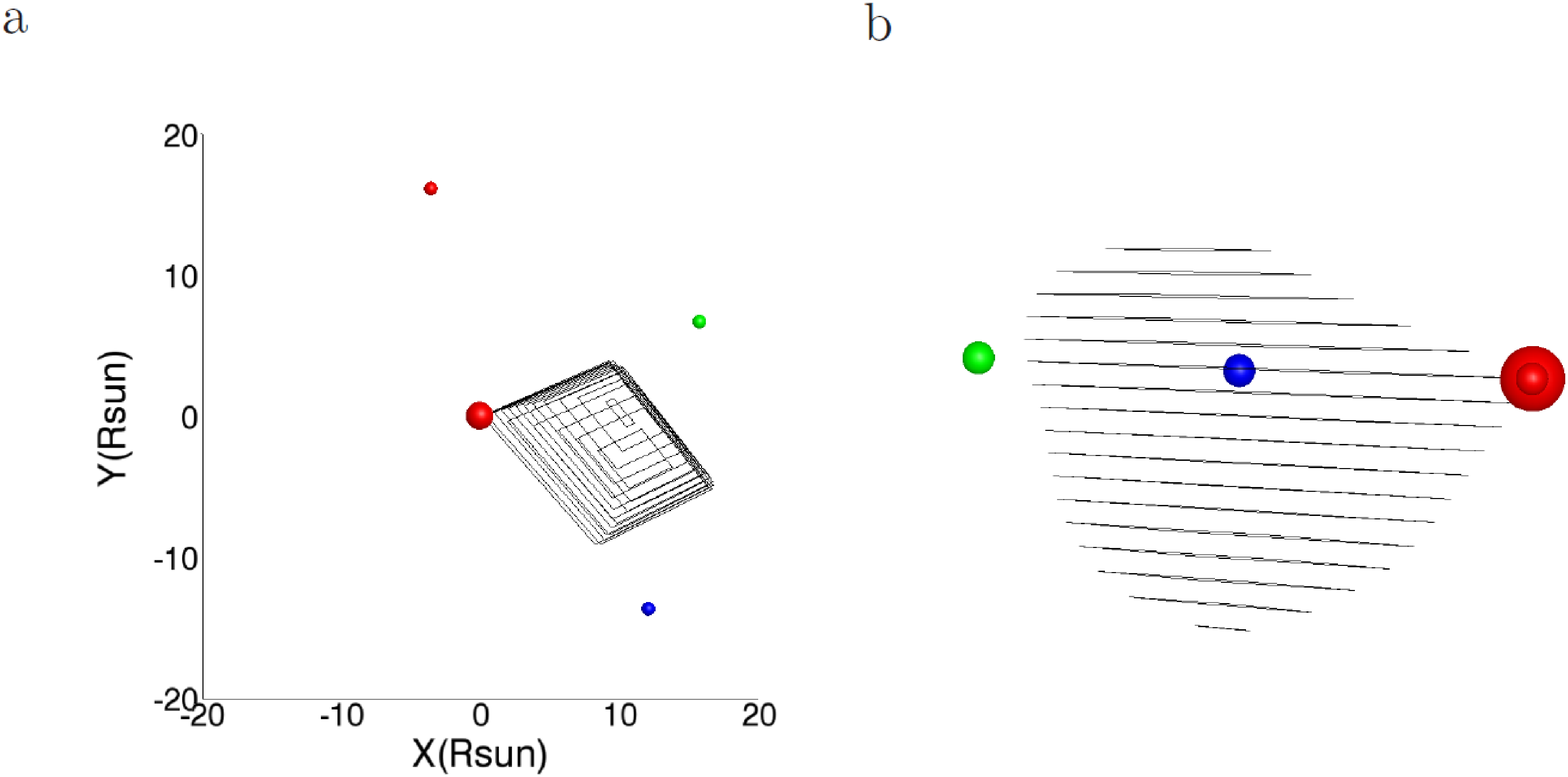}
  \includegraphics[width=12cm, height=5.cm]{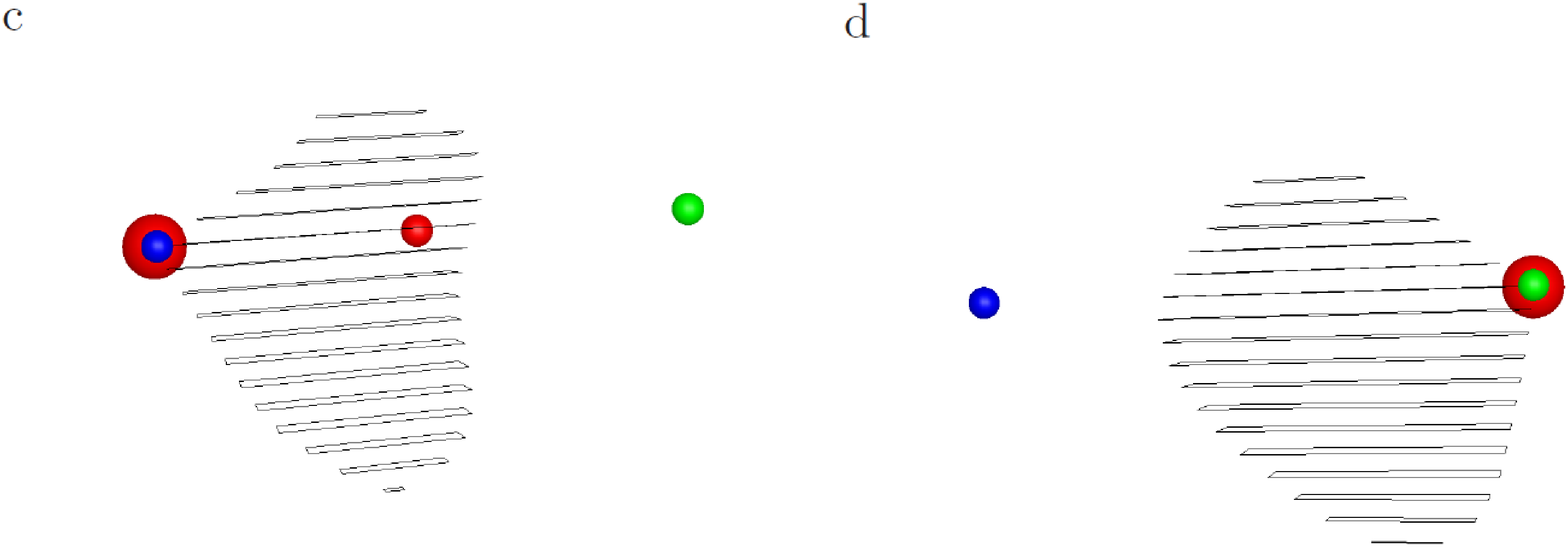}}
  \caption{Reconstructed 3D CME from SECCHI/COR2 B and LASCO/C3. Panel (a) 
  is a top view. The red sphere represents the position of the Sun. 
  Three smaller spheres from top to bottom indicate the positions of STEREO A (red), 
  SOHO (green) and STEREO B (blue), respectively. Their distances to the Sun 
  are scaled. Panel (b) to (d) are the reconstructed 3D CME as seen by STEREO A, 
  STEREO B, and SOHO.}
  \label{fig:geo_view}
\end{figure}

The geometric localisation method we use resembles the work in \inlinecite{deKoning:etal:2009}.
The reconstruction is based on the epipolar geometry \cite{Inhester:2006}. 
From two viewpoints
of spacecraft 1 and 2 at positions $\mathbf{r_1}$ and $\mathbf{r_2}$,
the epipolar reference axis is in the direction of $\mathbf{r_1}\times\mathbf{r_2}$.
For the stereoscopic reconstruction, all epipolar planes include the spacecraft
positions $\mathbf{r_1}$ and $\mathbf{r_2}$. The epipolar plane which also includes
the Sun center is normal to the reference axis. 
If one certain epipolar plane is chosen, its intersection with two image planes
yields two epipolar lines. 
In each image plane the intersection of the epipolar line with
the projected CME surface produces two intersections points, one on
the CME leading edge and the other on the rear edge.
The back projections of these four points form one quadrilateral in
each epipolar plane which contains the CME.
The stacked set of quadrilaterals on different epipolar planes yields a
volume which to a first approximation reconstructs the CME shape in 3D space. 

As an example, we present in Figure~\ref{fig:geo_view} the reconstruction 
that resulted from
only two views, SECCHI/COR2 B and LASCO/C3. Panel (a) is a view of the 3D CME from a vantage 
point above the north pole. From this particular viewpoint, the position of the 3D CME 
relative to the three view directions from STEREO A, B and SOHO can be easily
disentangled. For a low-latitude CME around the ecliptic plane, as the case in 
this paper, this information is very helpful 
to tell whether a CME arrives at the Earth or at another planet/spacecraft. 
A high-latitude CME may eventually deflect to the heliospheric 
current sheet close to the ecliptic plane where planets reside, e.g. 
\inlinecite{Zuccarello:etal:2012} and \inlinecite{Shen:etal:2011}.
Panel (b) to (d) are the projections of the 
reconstructed CME onto the image plane of STEREO A \& B and SOHO. They
have very similar shape to the CME observed by COR2 A and LASCO C3. However,
the CME shape in Panel (b) and in Figure~\ref{fig:cor_c3_2108} 
as seen from STEREO A do not
fit each other exactly. Since the reconstruction only includes the 
CME periphery from COR2 B and C3, it is not surprising that we arrive at this result.
The deviations reveal that observations from a third viewpoint will further
constrain the CME shape.

\begin{figure} % fig 3
  \centering
  \vbox{
  \includegraphics[width=6.5cm, height=6.5cm]{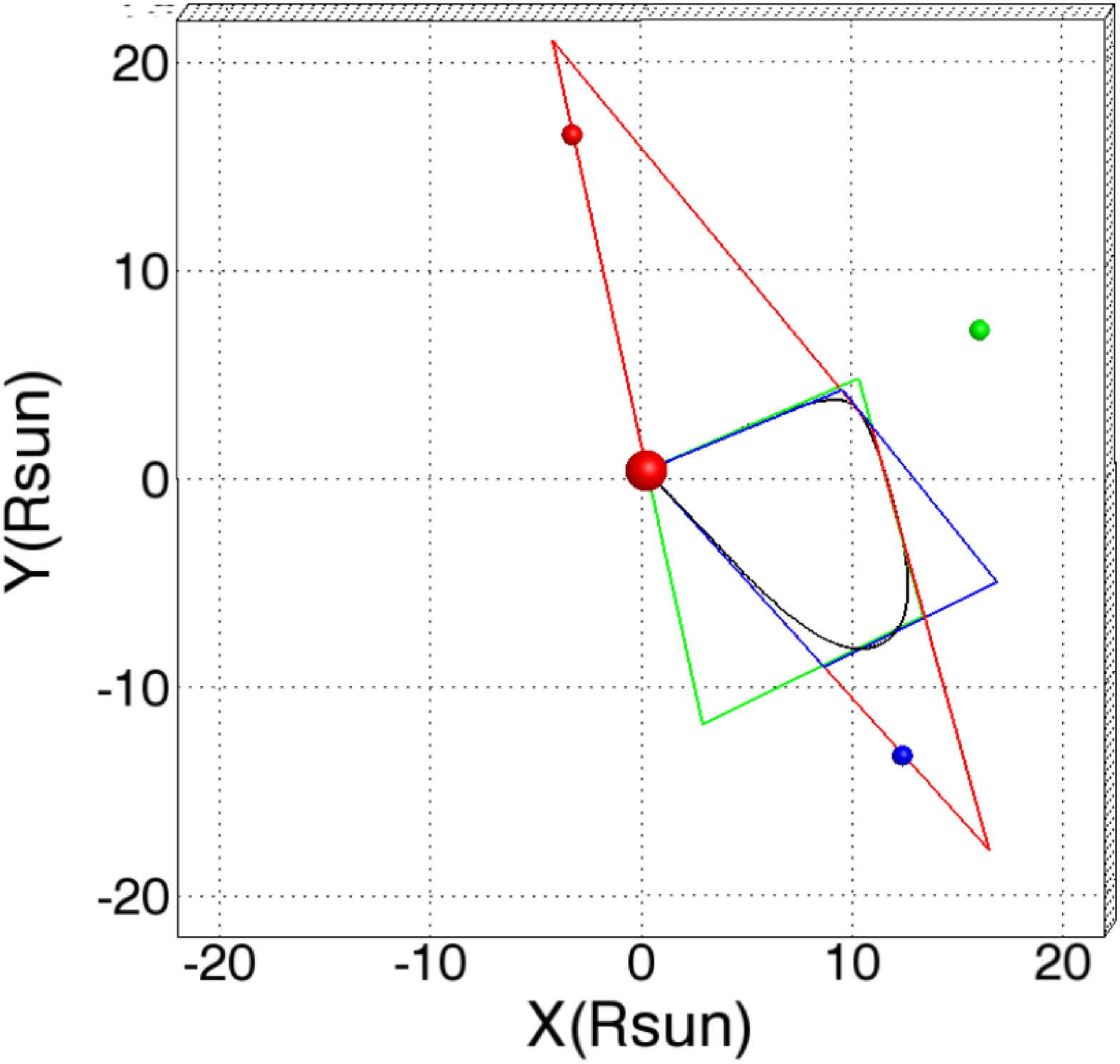}
  \includegraphics[width=6.5cm, height=6.5cm]{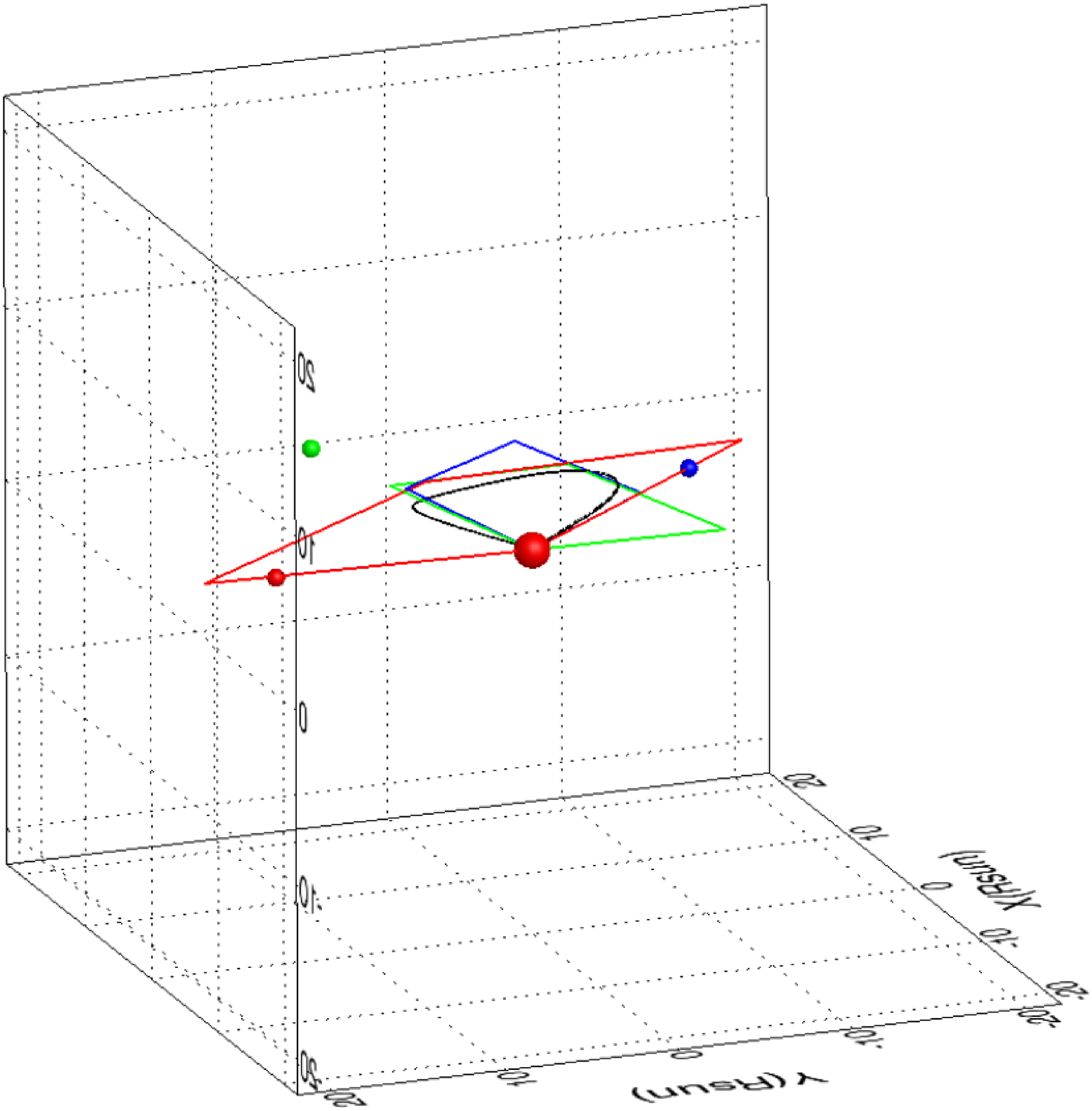}}
  \caption{Geometric localisation of the CME in the epipolar plane containing the
  solar centre from STEREO 
  A \& B in red, from STEREO A and SOHO in green and from STEREO B and SOHO
  in blue. The black curve is the reconstruction in the solar equatorial plane from the 
  mask fitting method in the next subsection. The red sphere represents the
  Sun. Three smaller spheres are STEREO A, SOHO and STEREO B.
  The color code of the three spacecraft is the same as in Fig. 2. Upper: a top view 
  of the reconstructions; Bottom: a side view showing the 
  small deviation of the epipolar planes from the equatorial plane.}
  \label{fig:geo_loc}
\end{figure}

In Figure~\ref{fig:geo_loc}, the red quadrilateral is the 
reconstruction from COR2 A and B in the epipolar plane across the solar centre. 
The green and blue quadrilaterals are similar to the red one, whereas constrained 
by COR2 A and C3, COR2 B and C3, respectively. Because three spacecraft were 
almost positioned in the same plane, these three epipolar planes (three quadrilaterals) 
determined by STEREO A, B, SOHO, and the Sun are roughly coplanar. 

It is obvious that the red quadrilateral extends in a much bigger range in longitude 
than the green and blue ones.
In another word, its corresponding uncertainty of reconstruction is much larger.
\inlinecite{Pizzo:Biesecker:2004} investigated the uncertainty 
of the 3D reconstructions with the geometric localisation technique as a 
function of the separation angle. Synthetic white-light image pair simulating 
STEREO-like coronagraph observation of a modeled 3D CME were created. The 
reconstructions from the synthetic image pair at different separation angles and 
the modeled 3D CME were compared. The ratio of the area of the reconstructed
quadrilateral to the actual area was around unity at a 90-degree separation, 
and increased almost monotonically to both 0 and 180-degree separations. 
In Figure~\ref{fig:geo_loc}, the reconstructions in red has an area much 
larger than the green and blue one due to the
large separation angle between STEREO A and B. This is consistent with the scenario depicted 
in \inlinecite{Pizzo:Biesecker:2004}.

Another immediate result which can be derived from Figure~\ref{fig:geo_loc} is 
that the common polygon area which results from the intersection of the red, green and 
blue quadrilaterals is more constraining than any of the individual 
quadrilaterals from just two view directions.
We therefore strongly recommend that the third viewpoint should be
included, if the data is available.
 
However, as mentioned in the beginning of this section, an epipolar plane depends 
on the position of the two observing spacecraft. If more than two spacecraft are 
used, epipolar planes still need to be constructed between each pair of spacecraft 
and epipolar planes from different spacecraft pairs in general will not match. 
However, if three spacecraft are located in a common plane with the Sun centre, 
the various epipolar planes can be approximated by planes parallel to the common 
spacecraft plane. This is equivalent to assuming an affine instead of a projective
viewing geometry. The error introduced is small \~O (r$_{CME}$/r$_{spacecraft}$) 
as long as the CME is observed close to the Sun, where the distance $r$ is measured
from the Sun. If the three spacecraft do not lie in a common plane the reconstruction by 
geometric localisation is still possible, but more complicated,
because it cannot be reduced any more to a set of planar intersections.
Instead the CME volume for each pair of spacecraft could be determined as a set of 
quadrilaterals on stacked epipolar planes and the intersections must be calculated 
from these volumes as a whole rather than from the quadrilaterals in each (approximate) 
epipolar plane.

\subsection{Mask fitting (MF)}

\begin{figure} %fig 4
  \centering
  %\setlength{\unitlength}{1cm}
  %\begin{picture}(12,12)
  %\put(0.25,6.5){\includegraphics[width=0.45\textwidth,clip=]{fig4a.eps}}
  %\put(5.8,6.75){\includegraphics[width=0.4\textwidth,clip=]{fig4b.eps}}
  %\put(0.2,0.5){\includegraphics[width=0.425\textwidth,clip=]{fig4c.eps}}
  %\put(6.25,0.0){\includegraphics[width=0.45\textwidth,clip=]{fig4d.eps}}
  %\put(0.,12){\makebox(0,0){a}}
  %\put(5.75,12){\makebox(0,0){b}}
  %\put(0.,5){\makebox(0,0){c}}
  %\put(6.,5){\makebox(0,0){d}}
  %\end{picture}
  \vbox{
  \includegraphics[width=12cm, height=5.5cm]{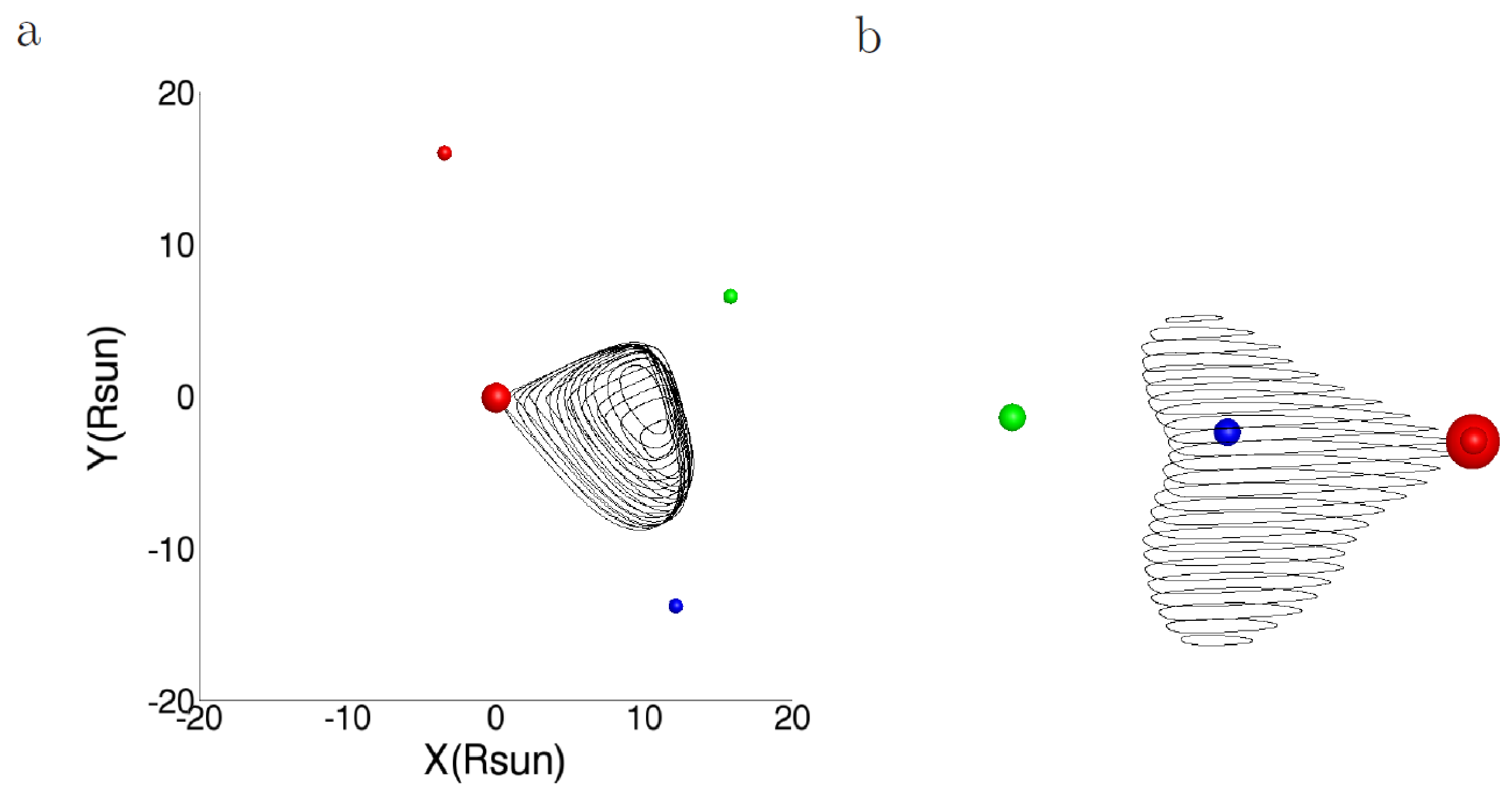}
  \includegraphics[width=12cm, height=5.5cm]{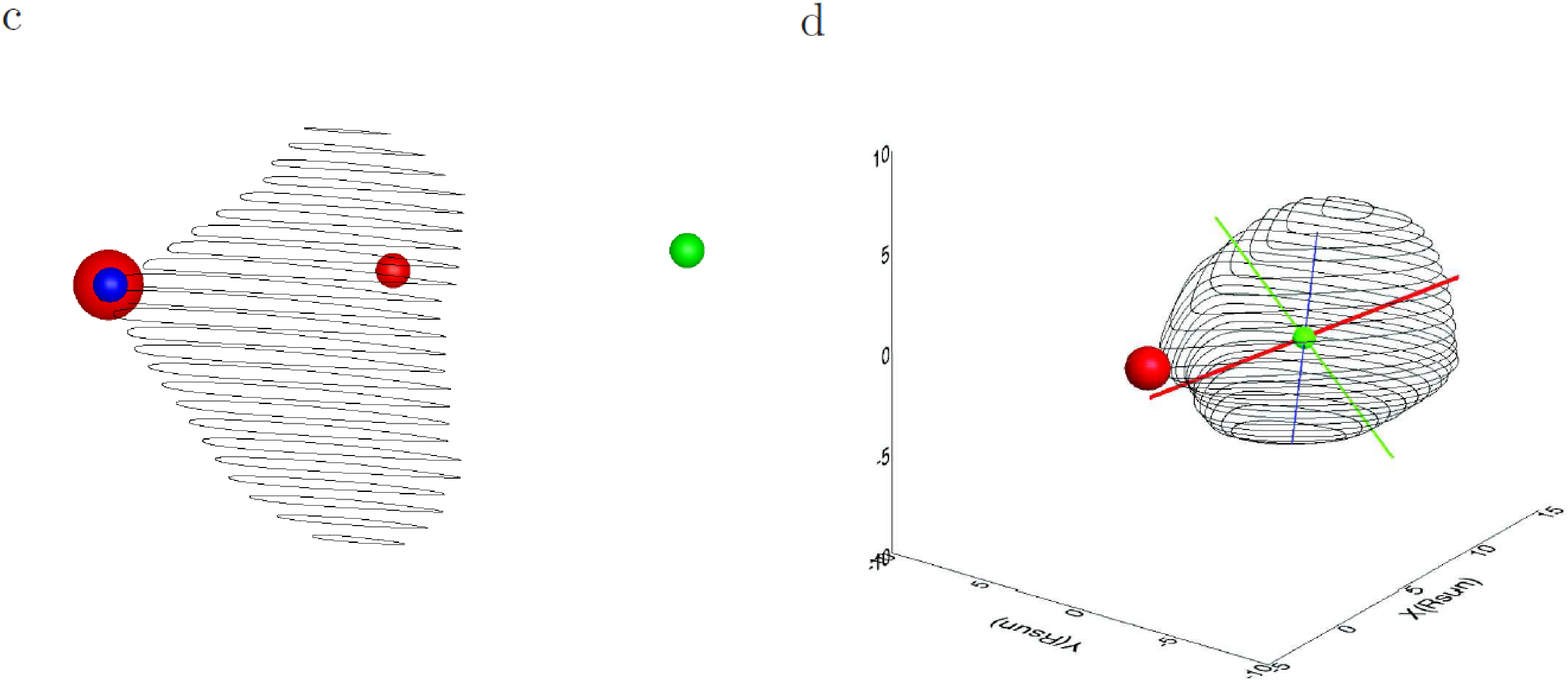}}
  \caption{Reconstructed 3D CME with the mask fitting method from SECCHI/COR2 B and LASCO/C3. 
  Panel (a) is a top view. The red sphere localise the position of the Sun. Three
  smaller spheres are STEREO A, SOHO and STEREO B. The color code of the three spacecraft is the same as in Fig. 2.
  Panel (b) and (c) are the reconstructed 3D CME as seen by STEREO A, 
  STEREO B. In panel (d) the 3D CME and its three principle axes are plotted. The
  three axes in red, green and blue are the major, intermediate and minor axes, respectively.
  The small green sphere represents the geometric centre of CME. }
  \label{fig:mask_view}
\end{figure}           

The mask fitting method geometrically follows up the previous concept, however,
instead of performing back-projections and finding intersections on epipolar planes, we use 
forward-projections from a discretised 3D space to verify whether a 3D point is potentially
inside or outside the CME volume. Using this concept, the epipolar geometry is not 
required explicitly. The details of this method are described in 
\inlinecite{Feng:etal:2012}. 

As a first step, three CME masks are created according to Figure~\ref{fig:cor_c3_2108}.
Pixel value is one inside the CME periphery and zero outside. 
Then we discretise a 3D cube centred at the Sun and in the range from 
-15 R$_{\odot}$ to 15 R$_{\odot}$ for all three coordinate axes. Each point
in this 3D cube is projected onto three image planes of STEREO A \&B and SOHO, respectively. 
Only those points which project into the masks of all three images are considered
to be the 3D points belonging to the CME. Afterwards, B\'{e}zier curves are 
employed to smooth the boundary of the resulting polygonal volume.
In Figure~\ref{fig:geo_loc} the black curve shows a slice of the smoothed reconstructed CME
volume in the solar equatorial plane z=0.

The different views of the reconstructed 3D CME with the mask fitting method can
be found in Figure~\ref{fig:mask_view}. Compared with the results obtained by using 
epipolar geometry explicitly with STEREO/COR B and LASCO data, the reconstruction
from three views provides a better solution, which can be verified by panel (b). 
The heart-like shape along the CME leading edge as seen from STEREO A is now 
clearly visible, whereas this feature is
not present in panel (b) of Figure~\ref{fig:geo_view}. In panel (d)
three principle axes and geometric centre of the 3D CME volume are shown.
Details of computation can be found in \inlinecite{Feng:etal:2012}.
Note that the geometric centre might not be the centre of gravity since a uniform 
density distribution was assumed. In this paper we use it as an approximation
of the centre of gravity to make comparisons with other methods.

\subsection{GCS forward modeling (FM)}

\begin{figure} % Fig 5
 \centering
 \includegraphics[width=12.cm, height=4.cm]{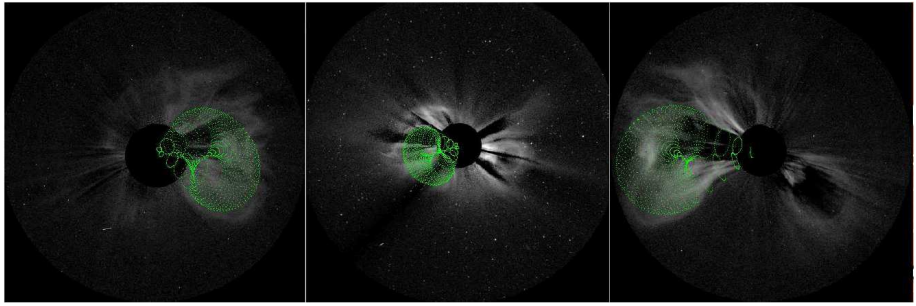}
 \caption{Fitting of a flux rope to the CME observed by COR2 A, C3 and COR2 B with
 the GCS forward modeling method.}
\label{fig:gcs_one}
\end{figure}

\begin{figure} % Fig 6
 \centering
 \vbox{
 \includegraphics[height=0.25\textheight]{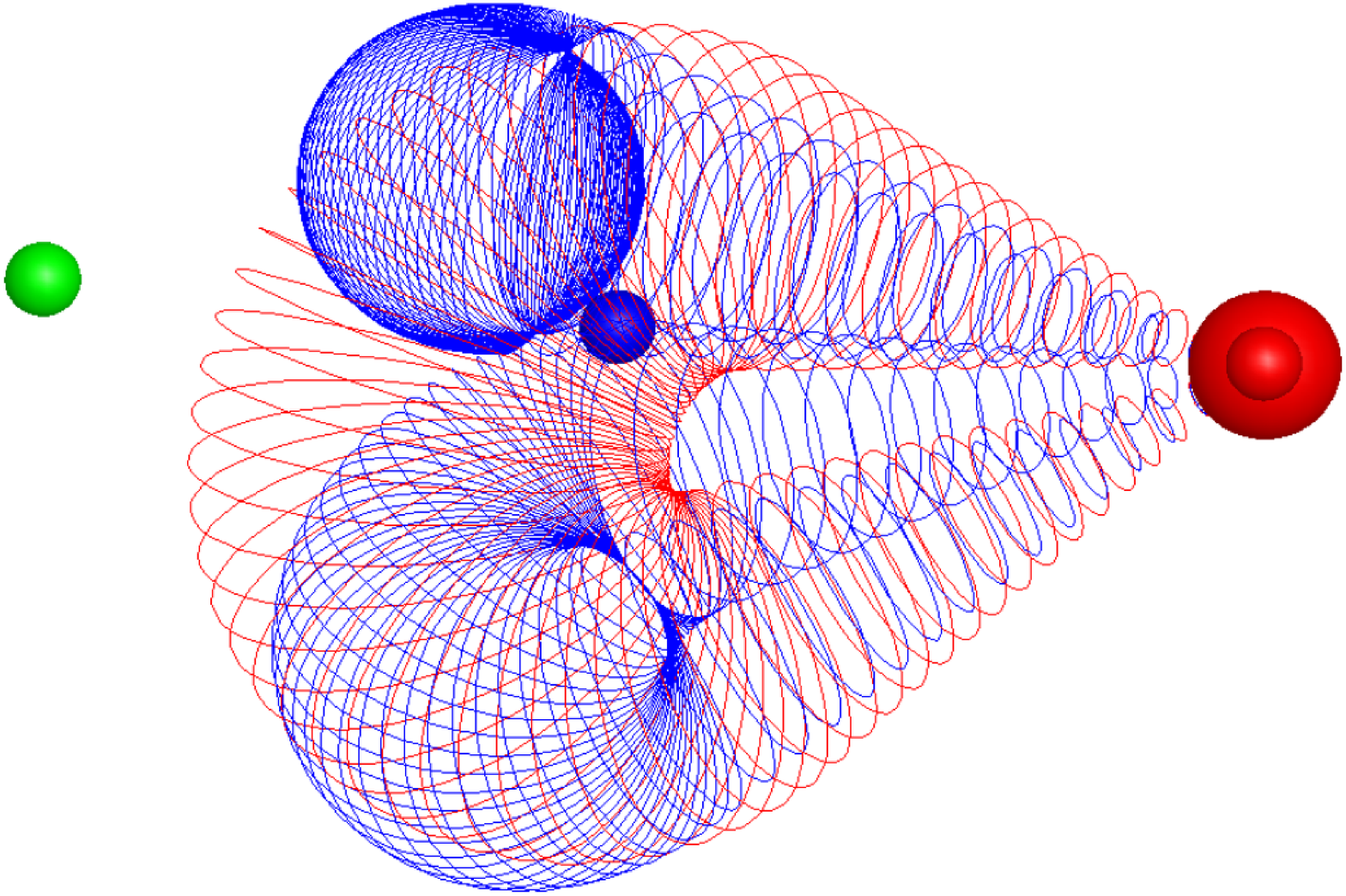}
 \includegraphics[height=0.25\textheight]{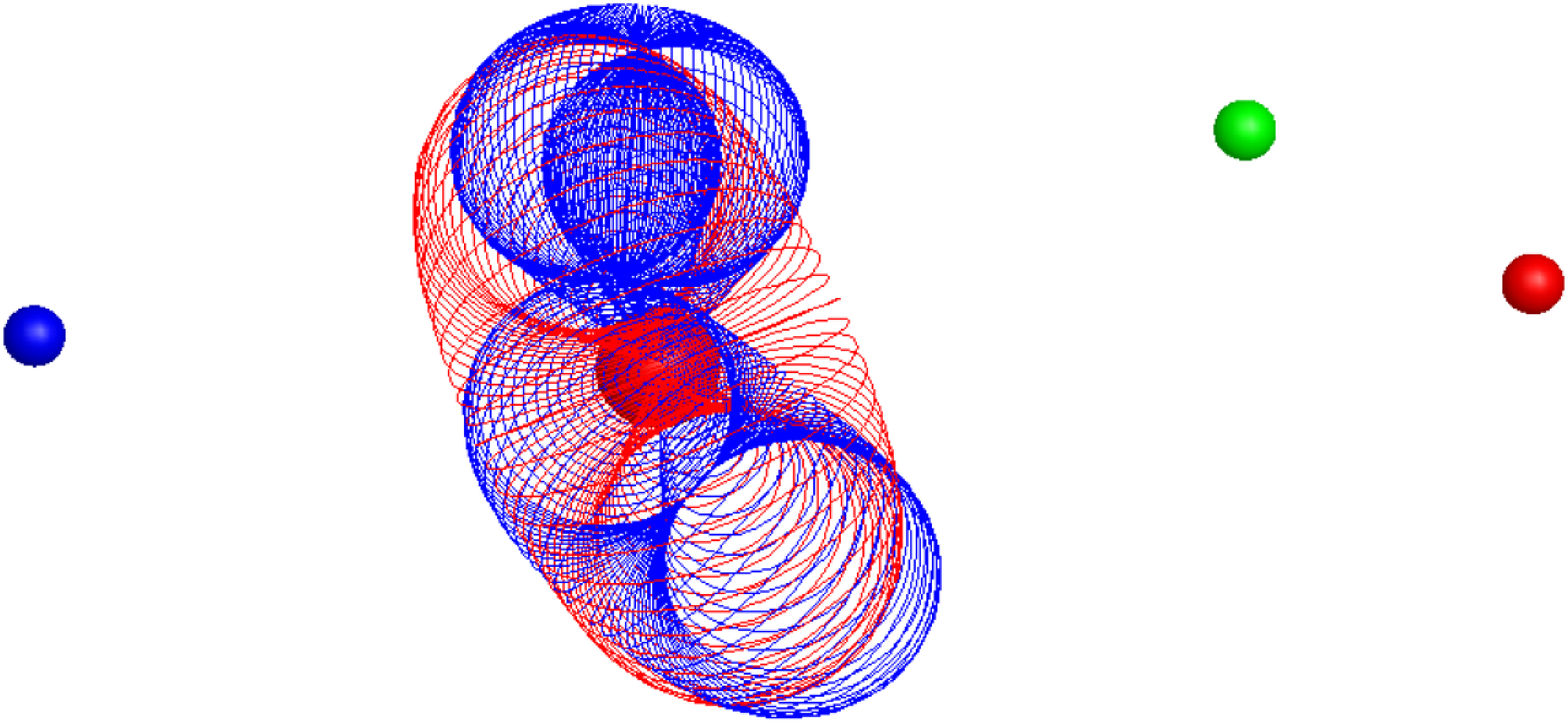}}
 \caption{3D flux ropes with the best-fit 
   parameters derived from the forward modeling method. The red flux rope 
   represents one single flux rope fitted to the CME observed from three 
   viewpoints. Two blue flux ropes are combined together to fit to the observed 
   CME. The red sphere stands for the Sun and three smaller spheres indicate the 
   directions to three spacecraft. The same color code as in Fig. 2 applies 
   here to the three spacecraft. Top: flux ropes as seen from STEREO A. Bottom:
   flux ropes as seen from the top of the fitted single flux rope.}
 \label{fig:gcs_two}
\end{figure}

\inlinecite{Thernisien:etal:2006} and \inlinecite{Thernisien:etal:2009} 
developed a forward-modeling technique for flux-rope-like CMEs using an 
empirically defined model of a flux rope, the graduated cylindrical shell (GCS).
It consists of a tubular section forming the main body of the structure attached
to two cones that correspond to the legs of the CMEs. The parameters constraining
the shape of the flux rope are determined by visual comparison of the projected
3D flux rope to observations. The optimisation is achieved by trial and error.
One hour after launch, the CME leading front became concavely deformed and its
shape can no longer match that of a single flux rope. Even though a
flux rope does a poor job for this CME, as a first estimate of the 3D propagation direction 
and the spatial extension of the CME, in Figure~\ref{fig:gcs_one} a flux rope is 
fitted to the observed CME from three viewpoints, i.e. STEREO B, LASCO and STEREO A. 
We followed the proposal of \inlinecite{Thernisien:etal:2009} and fitted two 
flux ropes to the observations from three viewpoints. The results are presented 
in Figure~\ref{fig:gcs_two} in blue color together with a single flux rope
in red for comparison.

\subsection{Polarisation ratio technique (PR)}

\begin{figure} % Fig 7
 \centering
 \vbox{
 \includegraphics[width=6cm, height=6cm]{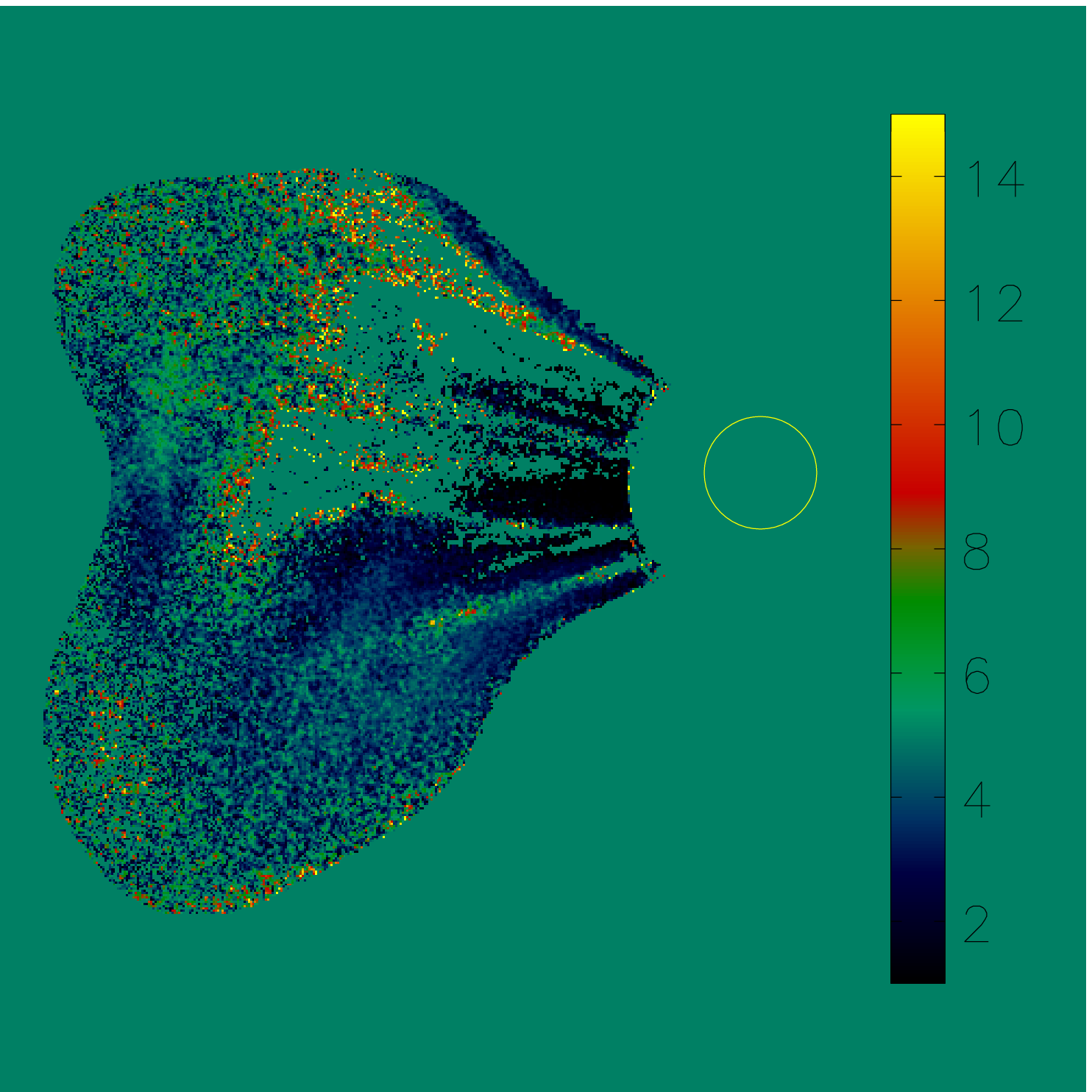}
 \includegraphics[width=6cm, height=6cm]{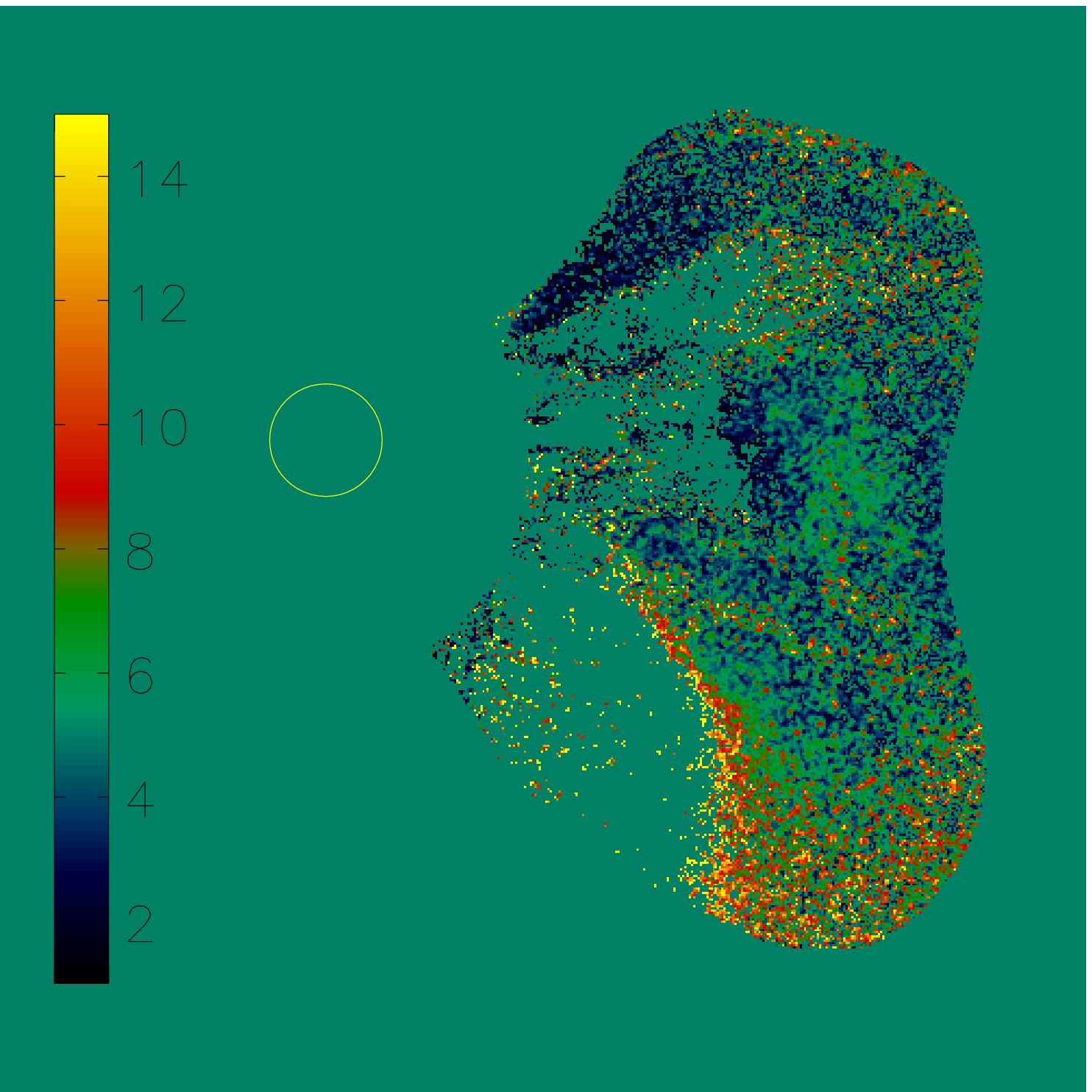}
 \includegraphics[width=8.5cm, height=8cm]{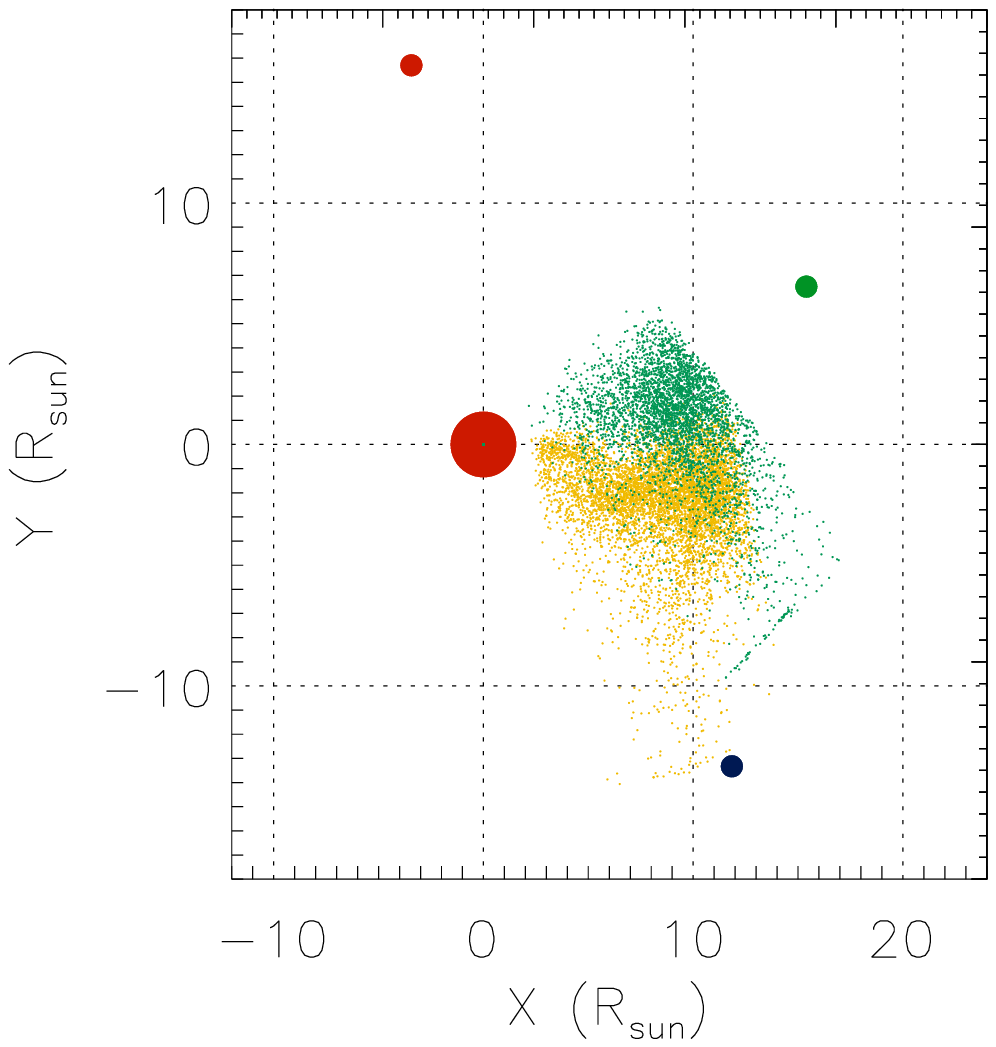}}
 \caption{Reconstructed 3D points from the polarisation ratio method. Top panels:
  the weighted distance of each pixel in the CME region to the respective POS calculated 
  from COR2 A (left) and COR2 B (right).The color bar scales the distance in 
  units of solar radius. The yellow circle mark the position of the solar disk.
  Bottom panel: A top view of the reconstructed 3D points. To make the dot
  density visible, we plotted the scattering centre every 50 points for all the 3D points.
  The yellow points are from the reconstruction from COR2 A, and the green points from COR2 B.
  The red sphere indicates the Sun, and three smaller ones mark the direction
  to three spacecraft. The color code of the three spacecraft is the same as in Fig. 2.}
 \label{fig:prab}
\end{figure}

The polarisation ratio technique was first applied to SOHO/LASCO coronagraph images
before the launch of STEREO \cite{Moran:Davila:2004,Dere:etal:2005}. Theoretically, the degree of 
polarisation of Thomson-scattered light by electrons in the corona is a sensitive function of
the scattering angle between the direction of the incident light and the direction
towards the observer \cite{Billings:1966}. Hence the ratio of the polarised brightness
$pB$ to the unpolarised brightness $tB-pB$, where $tB$ is the total brightness, is 
also a function of the scattering angle. Equivalently, this angle can be converted
to an effective distance of the scatterer from the plane of the sky (POS).
Observationally, the polarisation ratio can be retrieved from the polarimetric observation sequence
by LASCO or COR2 for each pixel within the CME region. Comparing with the theoretical
relation, we can derive a weighted distance of the centre of scattering along
the line-of-sight for each pixel.
\inlinecite{Mierla:etal:2009} applied this technique to several CMEs observed
by SECCHI/COR A \& B independently, and derived the distribution of LOS scattering 
centres of the CME. \inlinecite{Moran:etal:2010} has validated this method through triangulation. 
\inlinecite{DeKoning:Pizzo:2011} have suggested that this method 
can be used in space weather forecasting.

In order to remove the contributions (F-corona, stray-light, streamer contribution) 
other than the CME feature itself, a pre-event image was subtracted. The emission in the
coronagraph images is Thomson scattered light, but in some situations also Halpha emission 
can be observed (see e.g. \inlinecite{Mierla:etal:2011b}). If we assume that only Thomson 
scattered light was observed, then we could use the PR method to derive the 3D location 
of the CME. In order to increase the signal-to-noise ratio, the frames with the CME 
were smoothed by using a 5-pixel-by-5-pixel median filter. The size along LOS is 
restricted to 15 solar radii (in front and behind the POS), and the step 
along LOS is every 0.001 solar radii. The pixels where the intensity was smaller than 5\% 
of the CME region mean value were not included in calculation. This was done 
in order to reduce artifacts in the 3D reconstruction.

Fig. 7 shows the results of applying the PR method to the data of
COR2 A and B on the left and right panel, respectively. The color bar
indicates the distances in solar radii of the scattering centres off the
POS obtained for each image pixel inside the CME projection. The bottom
panel is a top view of the reconstructed 3D points. Yellow and green points 
represent the results from COR2 A and B, respectively. As the PR method can
only determine an absolute distance to the POS, an ambiguity of two symmetric
solutions arises. In cases that the CME source region is known, its position in
the solar disk can be utilised as a reference. From the close time of flare eruption and 
CME ejection. the source region is very probably the AR~11093. It was located on the 
western hemisphere as seen from 
the STEREO B. From STEREO A, it is not visible in front of the solar limb.
Therefore, we assume that all the reconstructed points from the PR method are behind
the POS as seen from STEREO A, whereas they are in front of the POS as seen from STEREO B.

Comparing the results from COR2 A and B in Figure~\ref{fig:prab}, we found that
in general they are consistent with each other. In the bottom panel of the top view, 
substantial part of the yellow and green points overlap. However, the misalignment is also 
obvious. The longitude of the centre of gravity differs by 19.4 degrees. 
In addition, the reconstruction from COR2 B in the top panels indicates that the southern part 
has a larger distance to the POS than the northern part, while the reconstruction
from COR2 A does not show this trend. We partly attributed the misalignment to the fact
that some features can only be seen by one instrument. In addition, in the above
reconstructions, we assumed that the CME is fully behind the solar limb as
seen from STEREO A and fully in front of the solar limb as seen from STEREO B.
In reality, this may not be true for all pixels. 

We note that 
two symmetric solutions exist for one pixel, i.e., one in front of the POS and the
other behind. This ambiguity can be partly disentangled by the identification of the
source region in EUV images or by the best match of stereoscopic observations.
However, CME source regions are not always apparent. e.g., for stealth 
CMEs \cite{Robbrecht:etal:2009}. Furthermore, without the information of the density
distribution along the line of sight, the determined distance to POS might
contain some errors \cite{Mierla:etal:2009} even if we include the stereoscopic results.
A further source of error of the polarisation ratio method may be that low temperature 
chromospheric
material might enter the coronagraph field of view whose radiation mechanism is
totally different from the Thomson scattering. Therefore, in those cases the polarisation
ratio technique does not work properly \cite{Mierla:etal:2011b}.
 
\subsection{Local correlation tracking plus triangulation (LCT-TR)}

\begin{figure} % Fig 8
 \centering
 \vbox{
 \includegraphics[width=9cm, height=9cm]{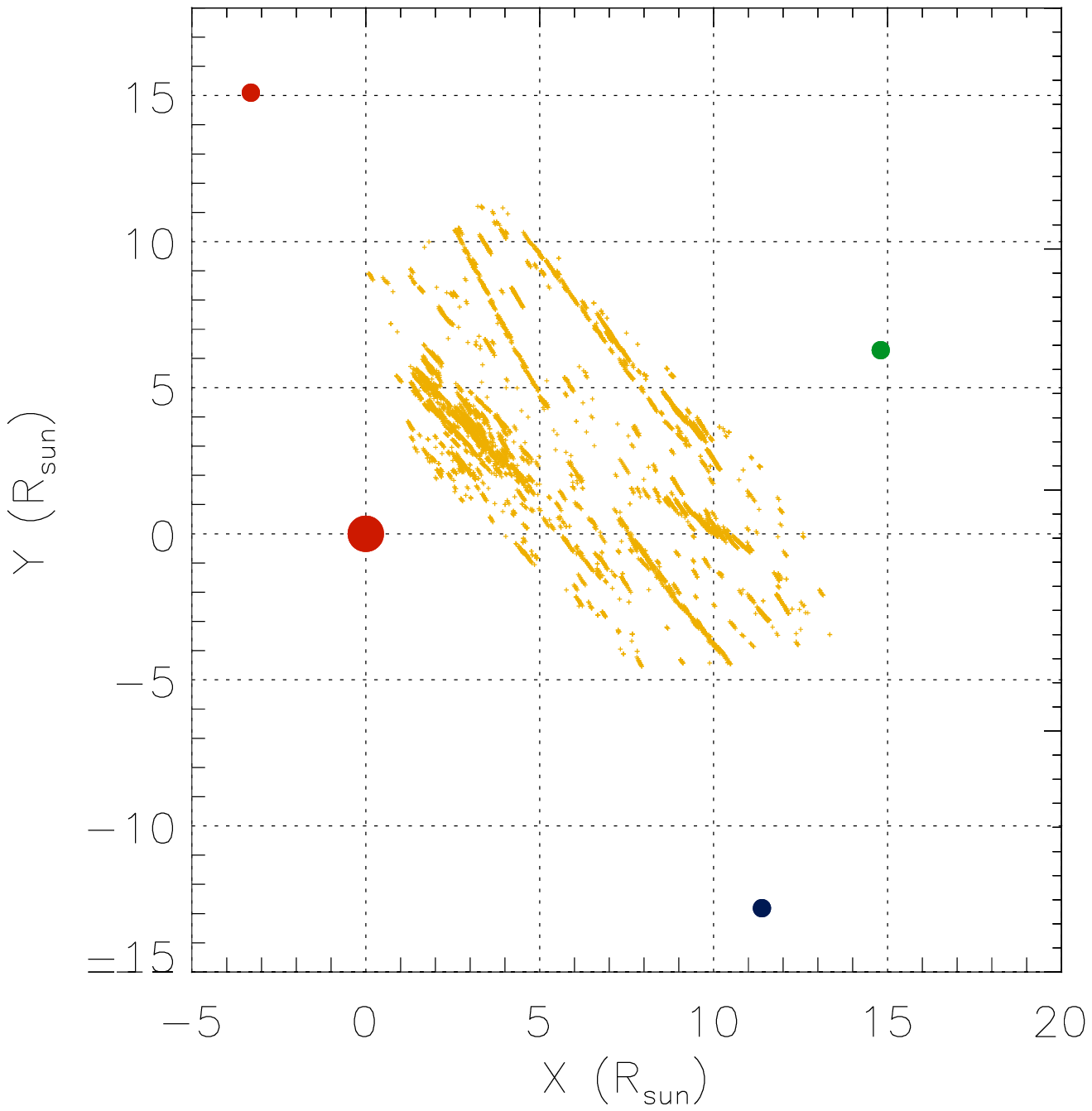}
 \includegraphics[width=11cm, height=7cm]{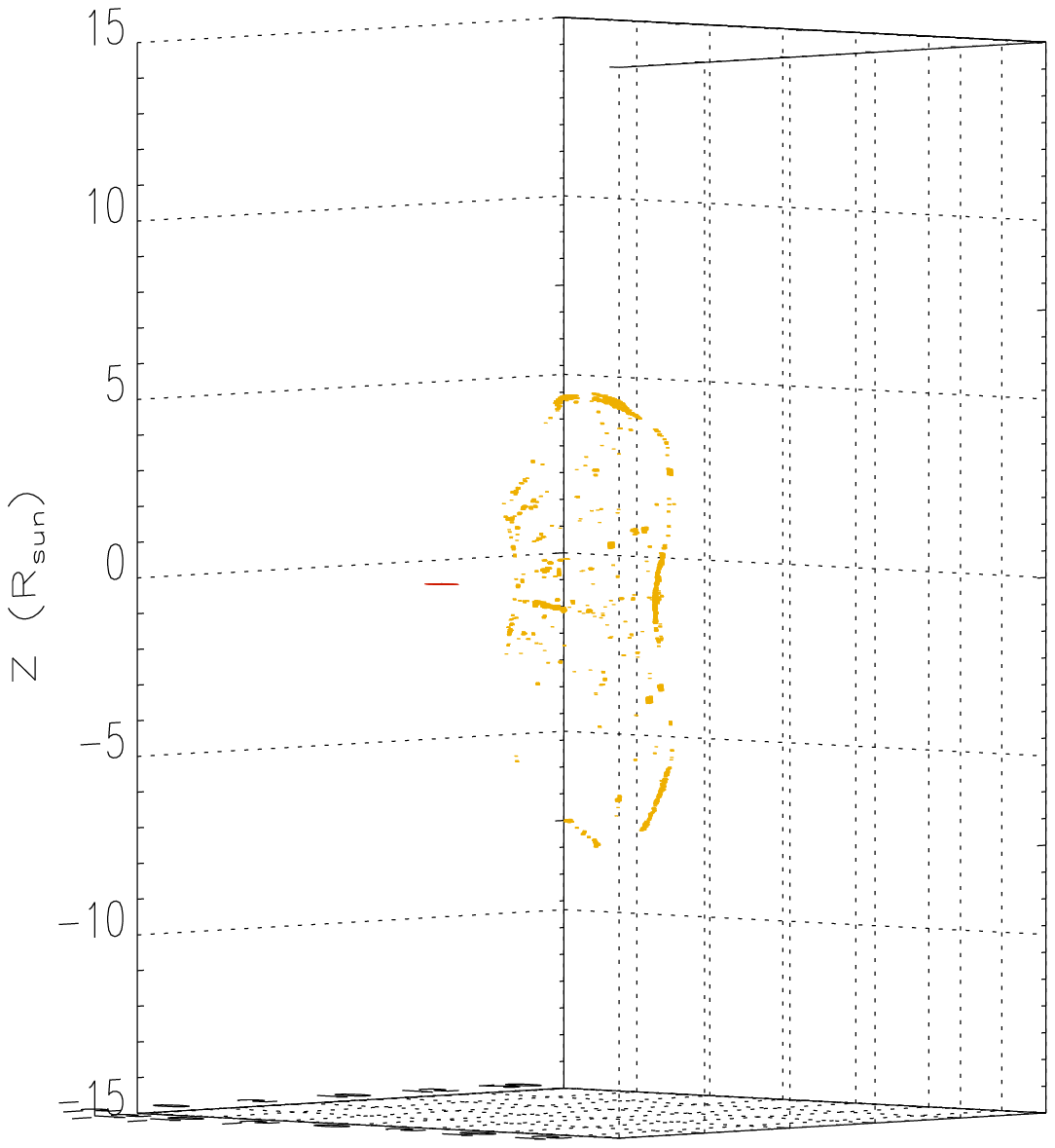}}
 \caption{Top panel: A top view of the reconstruction with the LCT-TR method 
 from COR 2 A \& B. The four spheres have the same meaning as Figure
 ~\ref{fig:prab}. Bottom panel: a view from STEREO B. In both panels, the 
 lower limit of the correlation coefficient is set to 0.7 for the correspondence. }
 \label{fig:lctab}
\end{figure}

A crucial step in this approach is to correlate the CME texture in both images before 
the triangulation. Since a CME often appears as a diffuse cloud, instead of the feature-based 
approach \inlinecite{Mierla:etal:2009} applied a local correlation tracking (LCT) method to 
find corresponding points in a STEREO image pair. In this method, 
we first co-aligned the images in 
STEREO mission plane such that the epipolar lines become nearly horizontal.
The elements to be matched are 
small subimages of a fixed size, called match windows. The criterion which decides whether 
two such windows in different images are positioned on the same object is the magnitude 
of their mutual correlation coefficient. The cross correlation is calculated from 
windows on two STEREO images on a common epipolar line. For a given position of one 
window on the epipolar line in one image, the corresponding position of the matching
window in the other image is determined by the correlation maximum.

Once the correspondence between pixels is established, the point in 3D space 
is calculated through back projection \cite{Inhester:2006}. In Figure~\ref{fig:lctab}, 
we present the LCT-TR reconstruction of the CME from COR2 data. 
A threshold of acceptance for the calculated correlation minimum of 0.7 was used in
Figure~\ref{fig:lctab}. In order to find correlations in two images, 
the match window (11x11 pixels) was fixed in a position on a given epipolar line 
in one image and move it along the same epipolar line in the other image. The search 
window was 512 pixels on the horizontal direction. As the CME was observed at the 
opposite limb in the two images, the search for corresponding features was flipped 
such that, for a pixel on the left side of image A, the search window was located 
on the right side of image B.
The upper panel is a top view. The reconstructed CME shows a propagation direction
towards SOHO which was, however, not validated by the in-situ observation close to the Earth 
\cite{Feng:etal:2012}. It may be that the lower acceptance limit of 0.7 for the correlation was
chosen too small causing a significant number of false correlations to be
passed. A higher value of the correlation limit, however, would have
reduced the number of correlations markedly.
The lower panel is a view from STEREO B. We found that most of the points within
the CME periphery have a correlation coefficient less than 0.7. The results of
LCT-TR clearly show a poor performance when the separation
angle of two STEREO spacecraft is large. 

\section{Comparison and Combination of different methods}

\subsection{Geometric localisation and mask fitting}
The reconstruction of the CME morphology is a typical inverse process
problem. The geometric locatisation method applies back projection to
find a solution whereas we use forward projections in the mask fitting algorithm.  
In Figure~\ref{fig:geo_loc}, we can see that the black curve from the mask 
fitting method fits the pentagon from the geometric localisation quite well. The small 
difference between the mask fitting and geometric localization is that the
epipolar plane does not necessarily lie on the solar equatorial plane of $z = 0$. They
deviate with each other by a few degrees as indicated in the bottom panel of 
Figure~\ref{fig:geo_loc}.
Nevertheless, from Figure~\ref{fig:geo_loc} we expect that in principle these 
two methods produce consistent results. 

Geometric localisation suffers from the limitation that the epipolar
plane is uniquely determined by the positions of the two spacecraft
and the point to be reconstructed. Only in the rare case that the
third spacecraft also lies in this epipolar plane the constraints
from the three viewpoints can be combined in a simple way. The mask
fitting method, on the other hand, can integrate the data from an
unlimited number of viewpoints with arbitrary positions. Obviously,
the more view positions are included in the reconstructions, the more
constrained and reliable is the result.

\subsection{GCS forward modeling and mask fitting}

\begin{figure} %fig 9
 \centering
 \hbox{
 \includegraphics[width=5.5cm, height=5.cm]{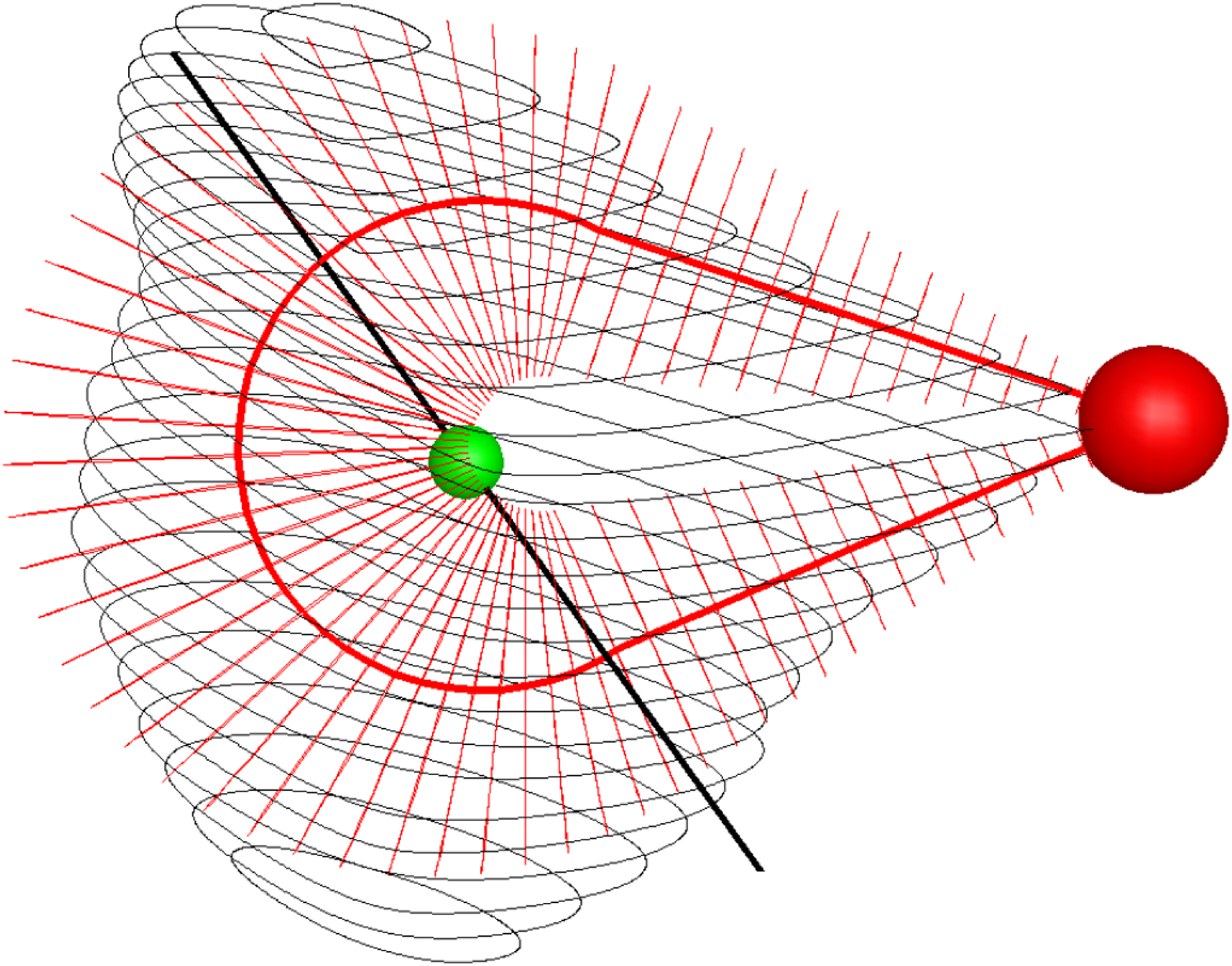}
 \includegraphics[width=5.5cm, height=5.cm]{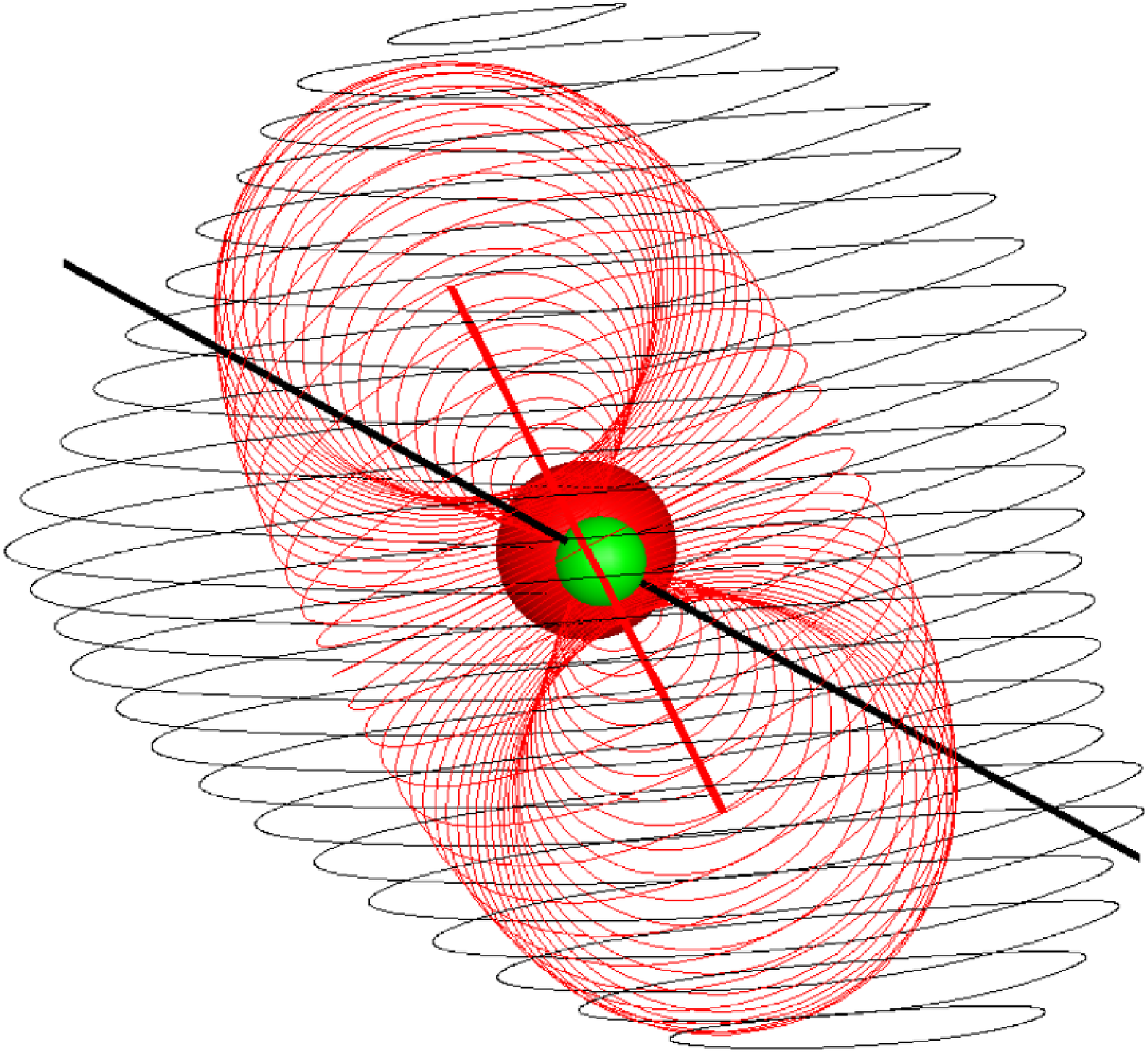}}
 \caption{Comparison of the single flux rope from forward modeling in red with
  the results from mask fitting method in black. The red sphere is the Sun.
  The red curve is the CME skeleton from the GCS model. The black straight line
  indicates the major principle axis of the 3D CME cloud derived with the mask
  fitting method. The front view of the flux rope is presented in the left 
  panel. The top view of the flux rope is in the right panel.}
 \label{fig:gcs_mask}
\end{figure}

As already stated in \S 3.3, when a CME does not have the shape of a flux rope,
the GCS forward modeling method can only provide an approximation of the 3D CME
morphology. It is the common feature for any forward modeling that the result
strongly depends on an a-priori geometrical assumption. However, real CMEs are not
always close to a flux-rope-like shape. Therefore, we found that the mask fitting
method has more flexibility than the forward modeling method. In the left panel of
Figure~\ref{fig:gcs_mask}, the fitted flux rope and the 3D CME reconstructed from
the mask fitting method are shown superposed along a view direction which is perpendicular
to the plane of the flux rope skeleton.
We can see how much the single idealised flux rope may deviate from the true CME.

Both the orientation of the flux rope characterised by the tilt angle 
$\gamma$ in \inlinecite{Thernisien:etal:2006} and the major principle
axis from the mask fitting method are in the northeast-southwest direction 
as can be seen in the right panel
of Figure~\ref{fig:gcs_mask}. The difference of the projection of the major principle
axis from the tilt angle from GCS model is about 30 degrees.

\subsection{Polarisation ratio and mask fitting}

\begin{figure} % Fig 10
 \centering
 \includegraphics[width=10cm, height=9.5cm]{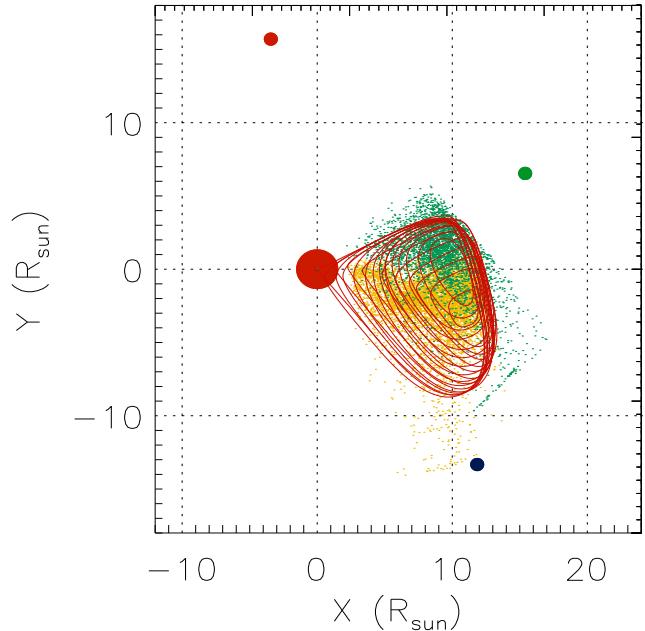}
 \caption{The top view of the 3D CME reconstructed with the MF technique in red
 color, and with the PR method from COR2 A \& B in yellow and green, respectively.}
 \label{fig:pr_mask}
\end{figure}

The polarisation ratio method has the advantage that only the data from one 
viewpoint and some additional information like the location
of the CME source region is required for a unique reconstruction. 
It yields some limited information also about the CME's
interior structure. However, the polarisation ratio can only provide the depth of a 
virtual scattering centre for each LOS but not the depth range over which the CME 
extends along the LOS. 
Figure~\ref{fig:pr_mask} shows an over-plot of the reconstructions with the 
polarisation ratio and mask fitting methods. Apparently, the MF method
locates the CME mostly in the overlap region of the results with the PR method 
from COR2 A and B images. 
 
Concerning the ambiguity of the two symmetric solutions from the PR method 
in \$ 3.4, we have resolved this
ambiguity by reference to the location of the CME source region.
In cases where the knowledge of the source region is not available, the CME
shape, reconstructed with the MF method could also be used as reference.
For the CME studied here, the MF result suggests that the CME centre
is completely in front of the POS as seen fom STEREO B and behind
the POS for most pixels of STEREO A. We therefore selected this 
respective solution for the PR method. We note that there is some probability
on lines-of-sight from STEREO A passing close to the Sun that the
PR scattering centre lies in front of the respective POS. Since the
MF method gives only the shape but no density distribution inside
the CME volume, predictions about the scattering centre must be treated
with care if the MF volume extends to both sides of the POS.
In addition, we found that the centres of gravity obtained with MF and PR A 
are quite close, however, those derived from MF and PR B differ by more 
than one solar radius. Similarly, \inlinecite{DeKoning:Pizzo:2011} combined the
3D CME reconstructed with geometric localisation method from two viewpoints
to resolve the PR ambiguity.

\subsection{Local correlation tracking plus triangulation and mask fitting}

\begin{figure} % Fig 11
 \centering
 \includegraphics[width=8cm, height=7.5cm]{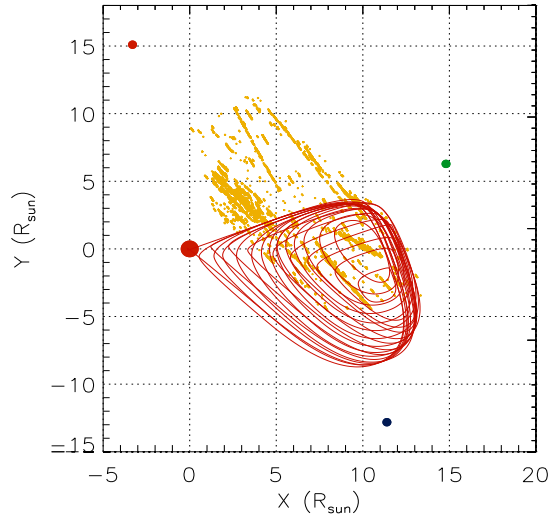}
 \caption{The comparison of the 3D CME reconstructed from MF and LCT-TR methods.
  The results are shown in red and yellow, respectively.}
 \label{fig:lct_mask}
\end{figure}

LCT-TR method works well for small separation angles \cite{Mierla:etal:2009}.
When the separation angle increases, the capability of LCT in solving the 
correspondence problem decreases. In Figure~\ref{fig:lct_mask} a top view of
the reconstructed CME with the MF and LCT-TR methods are indicated in red and 
yellow, respectively. We found a large discrepancy between them. The 3D CME 
location derived from the LCT-TR method indicates a propagation of the CME 
towards the Earth, whereas the CME position obtained with the MF method 
is consistent with a propagation towards the Venus \cite{Feng:etal:2012}. As already
pointed out in \S 3.5, the in-situ measurements around the Earth did not detect
any ICME related signature. However, the Venus Express spacecraft detected the shock and
ICME arrival \cite{Feng:etal:2012}. Therefore, we take this as an evidence
that the MF method for large spacecraft separation angles yields a more 
reliable 3D reconstruction than the LCT-TR method. The LCT-TR method 
appears to fit quite nicely inside the red quadrilateral obtained from geometric
localisation. In other words, LCT-TR suffers from the same large-angle
separation problem that two-spacecraft geometric localisaton does.

\section{Discussion and Conclusions}

In this study we made a comprehensive 
comparison of reconstruction results between five different methods, which can 
produce the morphological
shape of the 3D CME. The corresponding results and their comparisons with the newly
developed mask fitting method are presented as well. For a quantitative comparison,
we compiled in Table~\ref{tab:lonlat} the centre of gravity of the various 3D 
reconstructions and their range in longitude and latitude. From the fourth column
of the table, we list the maximum and minimum of longitude and latitude of the 
reconstructed 3D CME points. 

For the geometric localisation method, we have already found in \S 4.1 that its
result is consistent with the results obtained with mask fitting
method. Therefore, we present the longitude and latitude information for both methods
in the same row in Table~\ref{tab:lonlat}. For the forward modeling
method, the centre of gravity is obtained by assuming that the mass distribution
has the same formula as in \inlinecite{Thernisien:etal:2006}. As it is a symmetric
distribution, we expect that the centre of gravity lies in the axis of symmetry
of the modeled flux rope. Hence the centre of gravity has the same longitude and latitude 
as the source region central position. We found that the numerically calculated 
latitude and longitude widths of the fitted flux rope model is consistent with the 
values analytically given by Equation (1) and (2) in \inlinecite{Rodriguez:etal:2011}. 
For a single GCS flux rope, the analytical formula gives a longitude range of about 45 degrees and 
a latitude range of about 69 degrees. From Table~\ref{tab:lonlat}, these two figures 
which were calculated numerically are 46.6 and 69.6 degrees, respectively.

An inspection of the longitude and latitude of the center of gravity indicates that
MF/GL, FM and PR A result in similar values. The difference in longitude is not 
larger than 1.3 degrees, and in latitude is within 0.6 degrees. PR B shows a similar latitude,
however, the longitude deviates about 19 degrees. Even larger deviations from those of the 
MF/GL methods were obtained with the LCT-TR method, both in longitude and latitude.

Concerning the spatial extension of the CME in 3D, the results in Table~\ref{tab:lonlat} 
are not fully consistent with each other. For the FM method, a single flux rope inclined in the 
NE-SW direction is fitted to coronagraph images observed from three viewpoints. 
As mentioned in \S 3.3, the observed CME was strongly deformed away from a flux rope.
It is not surprising that forward modeling method has a different longitude and latitude extension than
the mask fitting method. Comparing with the MF method, we found a smaller longitude range
for the FM results. The latitude range derived from both methods agrees well.
Interestingly, the longitude range from the MF method happens to be in general 
consistent with the range of the overlap of the PR A and PR B results. 
There is a similar trend for the latitude minimum.
However, the latitude maximum of PR methods are 14 degrees higher. The longitude and latitude range
derived from the LCT-TR method shows a much bigger deviation from the other four methods. 
In cases of large separation angle, LCT-TR method does a poor job for the 3D reconstruction of CMEs.

In summary, for the event on August 7, we found that the mask fitting method 
achieves the same precision for the CME's 3D morphology as the geometric 
localisation method when the data from the same number of viewpoints are used. 
Compared with the forward modeling method, mask 
fitting is more flexible, no a-priori geometrical assumption is required. We found that
the mask fitting method can be utilised to remove the ambiguity in the polarisation
ratio method. In addition, the reconstructed CME with the mask fitting method
localises almost in the overlap region derived with the polarisation ratio method
from COR2 A and B. We did not find a consistent result of the local
correlation tracking plus triangulation with the other four methods.

In general, the mask fitting method produces a good 3D reconstruction of the CME 
surface if the data from three spacecraft with sufficient angular separation is
used. However, we are lacking the detailed internal structure
within the CME volume. In future, we will explore the MF method for the surface
reconstruction in combination with tomography to resolve the internal 
density distribution. The localisation of the CME relative to the plane of the sky is 
helpful to estimate the CME mass from coronagraph images, e.g., 
\inlinecite{Colaninno:Vourlidas:2009}. It can be compared to the mass calculation from
other observations, e.g., \inlinecite{Aschwanden:etal:2009} and \inlinecite{Tian:etal:2012}.
We anticipate that with
the information of the known propagation direction, spatial extension and density
distribution of a CME, a more precise prediction of space weather can be made. 

\begin{table}
\caption{Longitude,latitude of the centre of gravity, and longitude, latitude ranges 
of the reconstructed 3D CME with mask fitting/geometric localisation, forward modeling, polarisation 
ratio and local correlation tracking plus triangulation methods. The Carrington 
coordinate system is used. All the values are in units of degrees. }
\label{tab:comp}
\begin{tabular}{ccccccc}
  \hline                   % horizontal line
Method & lon.         & lat.         & lon. min         & lon. max     & lat. min      & lat. max \\
  \hline
MF/GL  & -13.3        & -7.7         &  -48.0           & 24.7         &-42.5          &25.3\\
FM\tabnote{derived from the fit of a single flux rope} 
      & -13.4         & -7.3         &  -35.6          & 11.0          &-42.0          &27.6\\
PR A  & -14.6         & -7.9         & -70.3           & 15.6          & -42.1         &39.1\\
PR B  & 4.8           & -6.8         & -40.3           & 41.7          & -55.6         &42.8\\
LCT-TR& 27.8          & 2.8          & -29.7           & 89.4          & -39.2         &39.2\\
  \hline
\end{tabular}

\label{tab:lonlat}
\end{table}

\acknowledgements
We thank S. Gissot for providing the LCT program.
STEREO is a project of NASA. The SECCHI data used here were produced by an
international consortium of the Naval Research Laboratory (USA),
Lockheed Martin Solar and Astrophysics Lab (USA), NASA Goddard Space
Flight Center (USA), Rutherford Appleton Laboratory (UK), University of
Birmingham (UK), Max-Planck-Institut for Solar System Research (Germany),
Centre Spatiale de Li\`ege (Belgium), Institut d'Optique Th\'eorique et
Applique\'e (France), Institut d'Astrophysique Spatiale (France).
LF is supported by National NSFC under grant 11003047 and by MSTC 
Program 2011CB811402. LF also acknowledges the Key Laboratory 
of Dark Matter and Space Astronomy, CAS, for finacial support. The contribution of BI 
benefitted from support of the German Space
Agency DLR and the German ministry of economy and technology
under contract 50 OC 0904. MM thanks MPS for financial support. Part of her 
work was also supported from the project TE 73/11.08.2010.
The work at the MPS was supported by DLR contract 50 OC 0904.

%% Bibliography
%
% Using BibTeX
%
%\bibliographystyle{spr-mp-sola}
% %\bibliographystyle{spr-mp-sola-cnd} %% Alternative style: no title, no concluding page
%\bibliography{cme.bib}
%
% Without BibTeX 

\end{article} 
\end{document}